\documentclass[12pt,journal]{IEEEtran}
\ifCLASSINFOpdf
  \usepackage[pdftex]{graphicx}
\else
\fi

\usepackage{subcaption}
\usepackage{amsmath}
\usepackage{graphicx} 
\usepackage{epsfig}
\usepackage{xcolor}
\usepackage{dblfloatfix}
\usepackage{url}
\usepackage[inline]{enumitem}
\usepackage[normalem]{ulem}
\usepackage{tabularx,booktabs}
\newcolumntype{C}{>{\centering\arraybackslash}X} 
\usepackage{lipsum}
\usepackage{booktabs}

\usepackage{xcolor,colortbl}
\usepackage{caption}
\usepackage{adjustbox}
\usepackage{hhline}
\usepackage{soul}
\usepackage{makecell}
\usepackage{hyperref} 

\begin{document}
\title{Effective Communications for 6G: \\Challenges and Opportunities}
\author{Ece Gelal Soyak, Ozgur Ercetin}

\setlength{\baselineskip}{5pt}
\setlength{\columnwidth}{220pt}

\maketitle

\renewcommand{\baselinestretch}{0.88}
\begin{abstract}
This article studies effective communication, one of the three forms identified by Weaver and Shannon, as an enabler for the upcoming 6G use cases.  The envisioned tactile, holographic, and multi-sensory communications require bandwidths in the order of terabits per second  and latencies in the order of microseconds for an immersive experience. We argue that a theoretical framework for transporting information tailored to end-users' goals is necessary to support such applications. Different from the recently emerging discussions focusing on the meaning of exchanged messages, we focus on using these messages to take actions in the desired way. We highlight the essential characteristics of distributed knowledge accumulation as a facilitator for this upcoming paradigm, and discuss the challenges of making effective communications a reality and the potential opportunities for future research to address these challenges. In a real-life use case, we showcase the potential reduction in the number of bits transferred owing to the transferred accumulated knowledge. 
\end{abstract}
\renewcommand{\baselinestretch}{1}

\section{Introduction}
\noindent In their seminal work, Shannon and Weaver \cite{shannon_weaver_1949} categorized communications into:\textit{ Level A: Technical}, \textit{Level B: Semantic}, and \textit{Level C: Effectiveness}.  Level A communication focuses on the accurate transmission of symbols. In contrast, Level B considers the precision of the conveyed meaning of the transmitted symbols, and Level C addresses the meaning of the symbols as it affects the receiver's conduct. Shannon addressed Level A type of communication with his classical information theory, and our Internet was born.  Historically, communication theorists did not examine Level B and C type communications since this would require considering  the philosophical content of the communication, such as limits of knowledge, emotion, and logic.  Nevertheless, Level B and C type communications remain an important research topic in  psychology, sociology, and cognitive sciences. 

In the next decade, with the envisioned 6G networks, novel applications will emerge utilizing holographic, tactile, and multi-sensory (i.e., visual, hearing, touch, smell, and taste) communications \cite{strinati20216g}. These new ways of communication will consume data from various sensors, demanding high bandwidth and permitting end-to-end latency of a few tens of milliseconds. Traditional methods of communication will soon be unable to support high bandwidth delay-sensitive global network traffic. 

To this end, we investigate an architecture critical to delivering only the {\em most informative} data such  that end-users can access information that is timely, useful, and valuable for achieving their goals. The paradigm shift from Level A-focused to Level B/C-oriented communication has recently been reinitiated by several works in the literature \cite{strinati20216g,seo2021semanticsnative,DBLP:journals/corr/abs-2201-01389,pappas,uysal2021semanticcomm,9530497}. Some of these discussions approach from a semantic perspective; the number of bits in the transfer is reduced by performing end-to-end semantic encoding \cite{strinati20216g,seo2021semanticsnative,DBLP:journals/corr/abs-2201-01389,9530497}. 
Others approach from mathematically well-understood Age-of-Information (AoI) related definitions, focusing on real-time applications such as monitoring and control \cite{uysal2021semanticcomm,pappas}. These assert that the information has the maximum effectiveness when it is fresh, and they propose prioritizing packets according to their freshness. However, for most 6G applications, even if latency is of paramount importance, one cannot reduce the effectiveness of the communication to a single dimension of latency. For example, effectiveness is closely tied to user perceptions in  multi-sensory interactive holograms. In this respect, emergent communication frameworks \cite{9466501,kim2019learning} appear as practical alternatives, where they use multi-agent reinforcement learning (MARL) to allow agents to develop their language to achieve their goals.  However, although these methods are flexible and applicable to most scenarios, their operation in an {\em end-to-end networking} setting has yet to be understood. Recently, a stochastic model of semantic-native communication inspired by human cognition and linguistics was proposed in \cite{seo2021semanticsnative}. The model uses a contextual reasoning module that allows the agents to obtain the communication context, assuming that the agents' backgrounds are known a priori.  However, in general, the agents' backgrounds are unknown at the beginning, and they also change over time.

This paper introduces the reader to the challenges and opportunities in effective (Level C) communication networks.  Effective communication deals with the way data affects the end-user's actions, whereas semantic (-native) communication is concerned with whether the data is interpreted correctly or not by the end-user. \textit{Hence, unlike semantic communications, effective communications take into account the time value of the data as it affects the receiver's actions.}  Of course, the receiver should interpret the meaning of the data before taking a possible action; however, the timeliness and the effect of this interpretation on the receiver's actions differentiate effective communication from semantic communication.  
In particular, quality of experience (QoE) is considered the crucial way to recognize users' satisfaction today. Effective communication captures the divergent QoE goals of applications that {\em act} upon their acquired knowledge while attempting to optimize the network-wide purpose of carrying fewer bits. 

\section{The Need for Effective Communications}
\noindent The effectiveness problem deals with the extent to which the meaning expressed (by the sender) and transmitted affects the receiver's conduct in the desired way.  Future networks enabling effective communications have the potential to trigger the desired effect by delivering as few bits as possible across the network.

\vspace{1mm}
{\bf Multi-sensory and holographic telepresence use case} --  Consider the necessary bitrate and maximum affordable delay of sensory communication to replicate the sense of touch at a remote location. The furthest receptors on the hand/feet from the brain are approximately 1.5 and 2 meters away, and the speed at which a sensory point transfers data is approximately 30 m/s; so the maximum delays of the signals are 0.038 s and 0.067 s \cite{9060950}.  A human can feel temperature differences with a resolution of 0.02$^\circ$C in the range $[5, 45]^\circ$C. Using this quantization to represent temperature data in 11 bits, and with sufficient sensory points, collecting temperature information $\sim$50 times per second, the bit rate for temperature information is  26.4 Mbps from the hands and 33.8 Mbps from the feet. Pressure sensors are more abundant in the human body; with a similar calculation, the required bit rates to carry pressure information from a palm, a fingertip, and a foot are 191.4, 21.2, and 116.6 Mbps, respectively. Thus, the total average bit rate to replicate the sense of touch on both hands and feet reaches 60.2 Mbps for temperature and 827.6 Mbps for pressure with a maximum delay of {\em a few tens of milliseconds}. Incidentally, when considering interacting with the hologram, the requirement increases to transfer multiple terabits per second with a maximum delay of 1 or a few milliseconds depending on the tactile operation \cite{9203885}.

{\bf Edge intelligence use case} -- This use case entails incorporating artificial intelligence (AI) to support seamless context-aware communications for controlling environments such as Industry 5.0, intelligent healthcare or autonomous transportation systems. In these applications, the sender and receiver are not in remote locations, they are collocated. The challenge is to support numerous (e.g. 10$^7$ connections per km$^2$) communications,  such as video uploads from multiple cameras on autonomous vehicles or the real-time health-monitoring data from (nano-)sensors. Naturally, a large number of connections requires high bandwidth. In addition, AI algorithms may run on heterogeneous platforms, necessitating low-latency communications for exchanging model parameters or arriving at consensus.

Note that in these examples, the goal is neither supporting a continuous stream of timely packets from sensors nor minimizing the mismatch between the actual and estimated processes of the user-received sensory signals. On the contrary, the  \textit{goal} is to make sure  {\em the system reacts swiftly to change}, e.g. person responding to change in temperature/surface tension, doctor responding to changes in remote patient's body, or autonomous factory responding to anomalies in machines' states. For the most effective use of network resources, the system design should aim to
provide and utilize context knowledge closer to the end users. 

\section{An Architecture for Effective Communications}
\noindent With human interactions,\textit{ effective communication} implies maximizing the impact on each other's behavior in an intended way. Similarly, our architecture facilitates delivering the necessary and timely information so that the message recipient takes the intended action. To design and analyze effective communication protocols, we ask the following fundamental questions:  1) How do we measure communication effectiveness via the time-varying actions of the receiver?  2) What is the information contained in the observed data, affecting the action of the receiver in the desired way? 3) How do we determine the shortest message to send to ensure the desired action is taken upon reception?  

Our design incorporates two fundamental features to answer these questions: i) A {\bf viability} metric to quantify the receiver's \textit{time-varying} level of satisfaction.  ii) The {\bf transfer entropy} metric to quantify the \textit{time-varying} information content affecting the viability.  We emphasize the \textit{dynamic} nature of the QoE and message content since the agents evolve during the process of observing the environment and actions. In the following, we present a high-level architecture that takes these factors into its design. The main actors in the envisioned architecture are data senders, receivers, and a hierarchical placement of evolving knowledge accumulators. We focus on these roles {\em per-application}, as every application has a distinct dynamic evolution in time, requiring different messaging forms and utilizing a specific language.

\subsection{Sending Shorter, Effective Messages}
\noindent For effective communication, the sender's role is to identify the next message that will most likely create the desired change of action in the receiver. Sender aims to send the shortest message from the application-specific language that prospectively maximizes the effectiveness at the receiver. For this, the sender maintains a {\em surrogate} of the receiver, using data from the observed network state as well as the in-application feedback (e.g., the receiver's head rotation in a Virtual Reality application). These observations help to assess effectiveness, which manifests as the receiver's QoE in that application. Such receiver-centric representation of QoE captures different end-user perceptions and allows customization per application and end user type.

We quantify effectiveness using a {\em viability function} representing the agent's ability to sustain and prolong its experience within the application. Using the surrogate receiver, the estimation of effectiveness  may be modeled using Deep Reinforcement Learning (DRL) that is trained using the history of previously sent messages and observed respective impact on the receiver. 

\subsection{Knowledge Accumulators (KA)}
\noindent KAs are application-specific neural networks trained using past experiences. Knowledge accumulation entails computing and storage; the presence of the latter alleviates the overhead on the former, particularly for compute-intensive edge intelligence applications.

\begin{figure}[tb]
    \centering
    \includegraphics[width=3.5in]{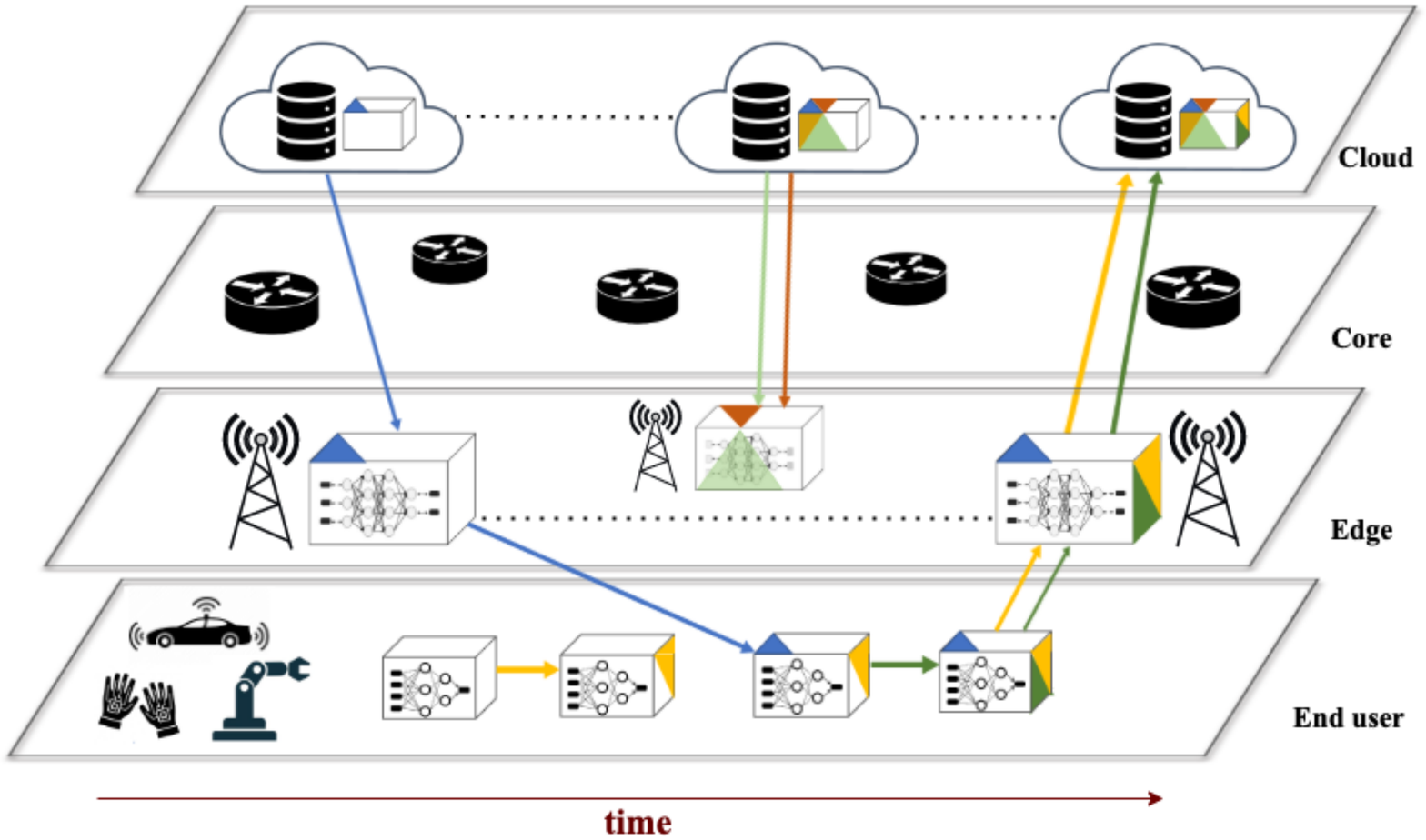}
    \caption{Distributed Knowledge Accumulation (KA).}
    \label{fig:KnowledgeAccumulation}\vspace{-2mm}
\end{figure}

Knowledge is dynamic; new experiences may render previous knowledge obsolete and sometimes wrong. Additionally, the accumulated knowledge in one application may have a different meaning in another or have different implications for other agents. As the entities in the network have unique experiences, knowledge accumulation should involve all stakeholders for the application-specific model to evolve fast and accurately. To address these concerns, our architecture comprises KAs placed at different vertical tiers in the network (on the end user, local and remote cloud). Evolution of knowledge (in time) and its distributed accumulation is depicted in Figure \ref{fig:KnowledgeAccumulation}.

Knowledge is accumulated at the receiver nodes, based on the incoming data and its {\em effectiveness}. The amount of knowledge that KA keeps and its validity duration are tightly related to accomplishing the application-specific goal. The accumulated knowledge accessible by a node is a function of its own experience and other receivers' accumulated learning using the same application.

Analogous to human communications, where  a speaker may use sophisticated terminology instead of long explanations when talking to an informed audience, the cognition of the accumulated knowledge available to the respective receiver would allow a sender to achieve a greater reduction in the number of bits transferred. 

\subsection{Interpreting Received Messages}
\noindent Upon receiving a message, a receiver attempts to maximize its utility by interpreting it using its accumulated knowledge. We formulate a receiver's conduct as a dynamic process that depends on its past actions and its accumulated (and inherited) knowledge. To identify dynamically acquired meaningful information, which we call {\em observed effective information}, we focus on a measure called {\em transfer entropy (TE)} \cite{transfer_entropy}. The key idea of TE is to assess the influence of the past value of a random process $Y$ (sender data delivered) on the current value of another random process $X$ (receiver action), when the  information regarding the past values of $X$ is accounted for. In this respect, TE represents the upper bound of increased communication effectiveness with the accumulated context knowledge. Note that in practice, latency and packet losses would postpone and even divert a model's training, reducing the actual improvement in the effectiveness. Such a reduction would be captured by the receiver's application-specific response, which would signal the sender to tune its optimization accordingly (e.g., send longer messages in this case) to approximate the effectiveness observable with TE.

\begin{figure}[tb]
\centering
\includegraphics[width=3.6in]{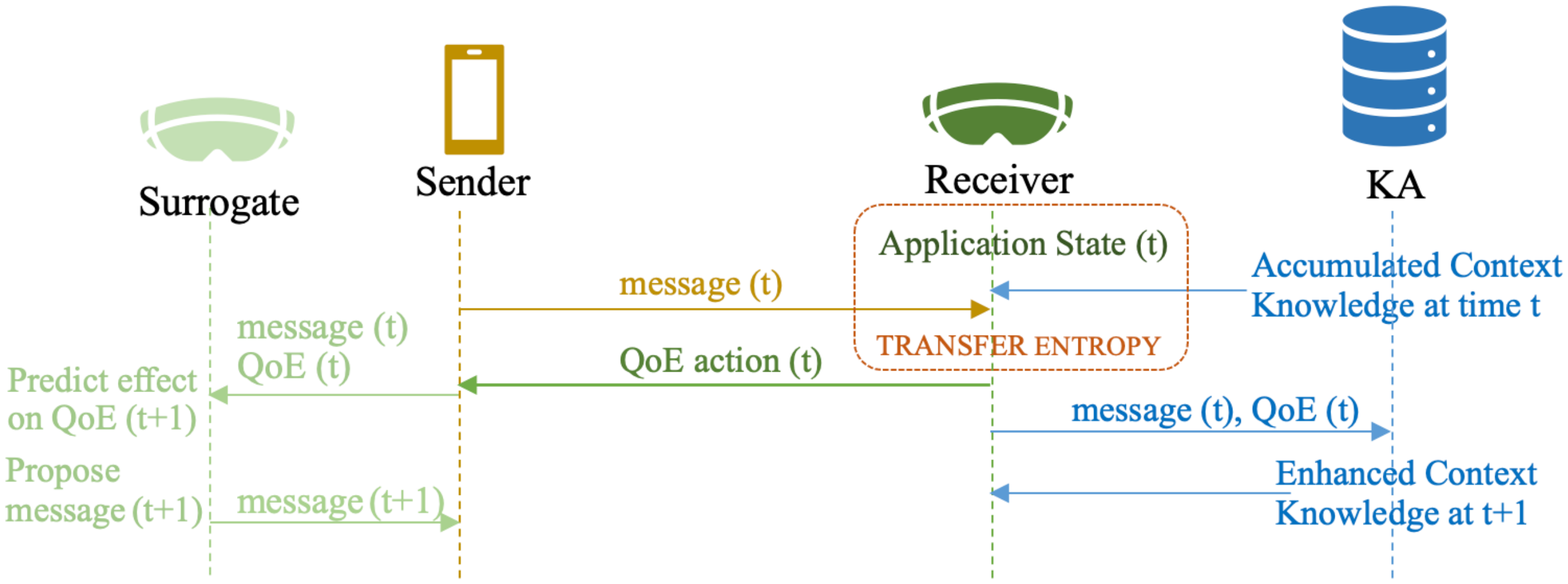}
\caption{Main steps in each iteration of the effective communications paradigm.}
\label{fig:CommunicationInstance}
\end{figure}

We explain the steps involved in an effective communication instance in Figure \ref{fig:CommunicationInstance}. The knowledge accumulated on the local KA at time $t$, the message received, the current state of the application impact the transfer entropy that assists the receiver in interpreting the sender's message. The receiver takes action within the application based on the impact on its QoE. The sender then determines its next message at $t$+$1$ based on its optimization with the surrogate receiver whose state is updated via the most recent exchange. Moreover, the state of the local KA evolves with the new experience at $t$, which is then updated on other network tiers (as depicted in Figure \ref{fig:KnowledgeAccumulation}). The implementation of each component involves a detailed investigation, regarding which we discuss the challenges and research opportunities next.

\section{Future Research Directions}
\label{sec:challenges}
\noindent Despite the recent interest in the so-called post-Shannon-era communications, the field is still in its infancy.  This section outlines research questions that need to be resolved before the potential of effective communication networking is realized. 

\vspace{-2mm}
\subsection{Effective Information Rate} 
\noindent {\bf Application-agnostic viability metric for measuring information} -- The agents communicating in an application aim to achieve specific goals. One may quantify the satisfaction of goals using a viability function indicating a system's self-maintenance \cite{InterfaceFocus}. As self-maintenance refers to the agent's ability to sustain and prolong its existence/experience, it is sensible to define the viability function to be application-specific; e.g., for plant control, the observable parameters are known, allowing more specific functions to be used \cite{uysal2021semanticcomm}. However, a universal and objective viability metric needs to be defined for the measurement of information. For example,  \cite{InterfaceFocus} proposes to measure information using the negative of the Shannon entropy since any other function, e.g., utility function, age of information (AoI), etc., will generally be scenario specific. Recently, information theorists  formulated semantic entropy relying on the availability of semantic information representation \cite{DBLP:journals/corr/abs-2201-01389}; meanwhile, the quest for quantifying semantic information content continues.

\vspace{2mm}
\noindent {\bf Rate of convergence to the optimal dictionary} -- The objective of the communication between agents is to achieve the maximum \textit{viability} of the receiver by carrying the minimum possible number of bits over the network. The (viability-) optimal information transfer will achieve the exact value of the viability metric as that for all bits transferred; with redundancy removed, the achieved viability may be more significant. The goal is to find that viability-optimal information transfer that yields the minimum transfer entropy. This study should consider the fundamental tradeoff between the costs of computation and communication.

\vspace{2mm}
\noindent {\bf Estimation of entropy from empirical data} -- The proposed way of measuring the effective information rate  requires the entropy to be calculated from the observed data.  However, the empirical calculation of entropy is challenging since the simple use of formulae of entropy and mutual information, with the probabilities estimated, suffers from bias, and it has finite variance.  Additionally, in many cases, the size of the dataset over which the entropy is calculated would be small. In that regard, a feedback mechanism from the end-user to understand the entropy statistics may improve the estimation. However, this requires additional complexity and communication costs, and one should weigh its gains against costs. Finally, although TE is well studied, and several numerical estimation methods are available for evaluation, these algorithms only work in hindsight, i.e., after an action is taken for a given observation \cite{DBLP:journals/corr/abs-1912-07277}. 

\subsection{Emergent Knowledge Accumulation}  
\noindent {\bf Evolution of a shared language of transferred information} -- Effective communication requires coordination among the agents.  Two (or more) agents need to establish and use a purpose-built dictionary that would help them maximize their mutual benefits from their communication. Ideally, this adaptation should be autonomous, i.e., the change of parameters in data exchange should be decided without even knowing the application or the agents' backgrounds. This objective fits well in the field of \textit{emergent communications}   that studies which circumstances lead to communication as an instrumental strategy when multiple learning agents are rewarded for completing specific tasks  \cite{10.1162/isal_a_00399}.

\vspace{2mm}
\noindent {\bf Knowledge distillation} -- With advancements in federated learning techniques, knowledge can be aggregated as individually trained models are integrated. Proprietary applications may adopt this paradigm earlier and potentially generate and deploy trained models to minimize variations in end-user QoE. In addition, DRL models trained on large datasets within the scope of an application may be adopted, using {\em heterogeneous transfer learning}, to be used in the context of other applications \cite{TransferLearningSurvey}.

\subsection{Locating Knowledge}   
\noindent {\bf Optimizing KA placement} --  KAs should be close to the data consumer (e.g., on end users or edge), to provide information with low latency  and to project communication overhead on fewer links. On the other hand, aggregating knowledge into a more extensive scope (e.g., in the network core or the cloud)  facilitates better QoE with more communications. This tradeoff must be studied for determining the optimal locations of KAs given {\em application specific requirements}.

\vspace{2mm}
\noindent {\bf KA service discovery} -- The agents need to locate and access the {\em knowledge} distributed in the network. This functionality requires an efficient way to name these services and resolve names for supporting end-user applications.  Research on Named-Function Networking (NFN) and Information-Centric Networking (ICN) may guide in addressing these problems. However, there are still challenges as both the user background and knowledge are dynamic; unlike in NFN, the effect of  KA at the receiver depends on the time of the request and the identity of the requester.

\subsection{Effective Networking Protocols}
\noindent 
{\bf Fairness for effective communications} -- The functionality in Figure \ref{fig:CommunicationInstance} presents an overlay on top of the traditional layered network architecture. While this is practically more achievable, the adaptation of effective communications paves the way for {\em a paradigm shift in networking protocols}. As an example, {\em fairness in effective communication} needs a new interpretation, e.g., the bottleneck link of effective communication must be shared to ensure all flows over that link deviate from their ideal \textit{action} by the same amount. Medium-access protocols should also be redesigned to incorporate the viability of end-users in transmission scheduling.

\vspace{2mm}
\noindent {\bf Reliability of knowledge transfer} -- For traditional (Level A) communications, data transmitted by the source must be perfectly reconstructed at the receiver. However, relying on past experiences and the application context, a receiver could still take the correct action even though the received message is distorted. With effective communication, contextual reasoning would eliminate the need for redundancy. This motivates redesigning data link and transport protocols. Note that the computational needs of effective communications are higher than those of traditional communications; effective network protocol design should aim to balance the computational and communication overheads.

In addition, future network protocols should be able to support the agents to {\em learn} the communication protocol(s) to adapt and use. There are a few recent works that address \textit{learning-to-communicate} over noisy channels \cite{9466501}, and over a common broadcast medium \cite{kim2019learning}. Reward-shaping methods, such as those used in multiagent reinforcement learning methods, may be used to expedite the learning of protocols.

\subsection{Economic Aspects} 
\noindent {\bf Challenges for network operation} -- Economic gains from effective communications will be a deciding factor in their development. For one, effective communications will facilitate achieving desirable end-user QoE in the face of new applications' requirements. However, knowledge accumulation entails  computationally intensive deep learning to identify useful information for the end-users. Trained models also incur a storage overhead on different network components.  Intent-based networking (IBN) solutions may be used for identifying agents limiting the performance and allocating resources to enable effective communications.

\noindent {\bf Need for incentive mechanisms} -- Incentive mechanisms are needed to induce cooperation 
whenever knowledge accumulation is non-altruistic. For the end-user, there is a tradeoff between the QoE and the overhead of knowledge accumulation; the latter is computed via additional local resources or retrieved from external accumulators. For the operator, in-network processing may yield a user churn versus cost tradeoff. Joint and complementary  incentives for cooperation between end-users, network operators, and application service providers will be necessary to engage all stakeholders. 

\subsection{Privacy and Trust}  
\noindent {\bf Tradeoffs between efficiency and privacy} -- The effect achieved in communication is proportional to the knowledge accumulated to assist interpreting received messages. For the KA models to quickly evolve to a state where the desired effect is achieved via  shorter  messages, it is necessary to train more comprehensive models with many users' traffic in a more extensive scope. However, some data may not be suitable for sharing. For example, it may contain personal identifiers such as location, time zone, device type, user behavior,  or application context (e.g., remote surgery or AR-assisted collaboration may reveal user data or intellectual property). Hence, there is a  tradeoff between privacy and effectiveness, i.e., involving a distributed, multi-tier knowledge accumulation mechanism may increase the QoE of the end-user but may jeopardize its privacy. 

\noindent {\bf Trust mechanisms} -- 
The effectiveness of the communication is also based on how much the user and the external KA (i.e., deployed on the edge or cloud) trust each other. In particular, mechanisms of the end-user should discount any information received from a partially trusted edge or cloud KA when taking action. Likewise, the KAs deployed on the edge or cloud should not integrate the knowledge accumulated by untrusted agents. Distributed trust mechanisms such as one using blockchains may be designed for validating contributions from KAs in either direction. 

\section{Communicating Effectively for Mobile User Handover:  A Use Case}
\noindent In this section, we demonstrate the gains that effective communications can provide over conventional (Level A) communications.  For this purpose, we analyze the user equipment (UE) handover between the base stations (BS) in LTE networks.  

\begin{figure*}[!t]
\centering
\begin{subfigure}{0.9\columnwidth}
    \includegraphics[width=3.1in]{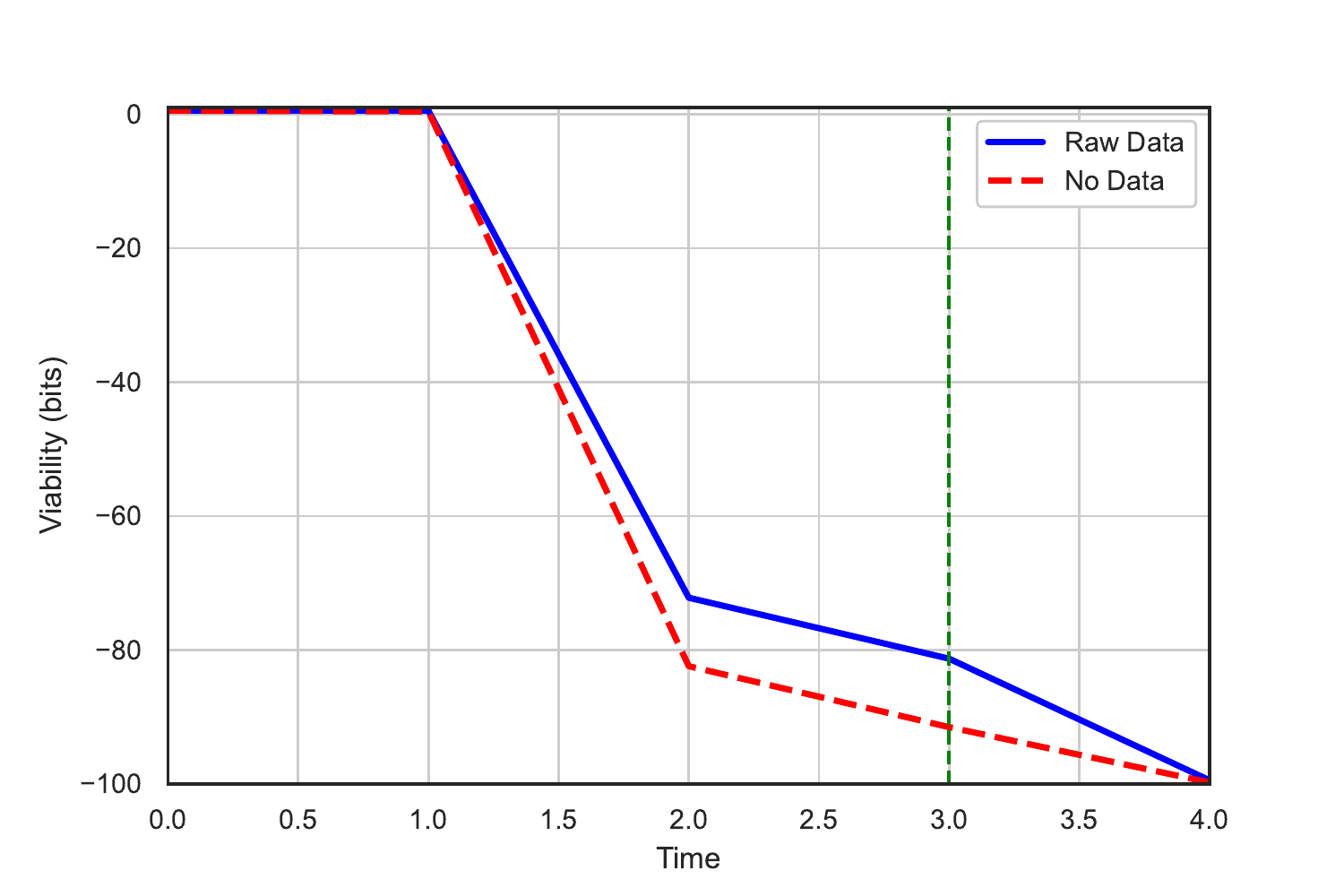}
    \caption{Viability with respect to time for a mobile UE with and without measurement data.}
    \label{fig:viability_1}
    \end{subfigure} \,\ 
\begin{subfigure}{0.9\columnwidth}
    \includegraphics[width=3.1in]{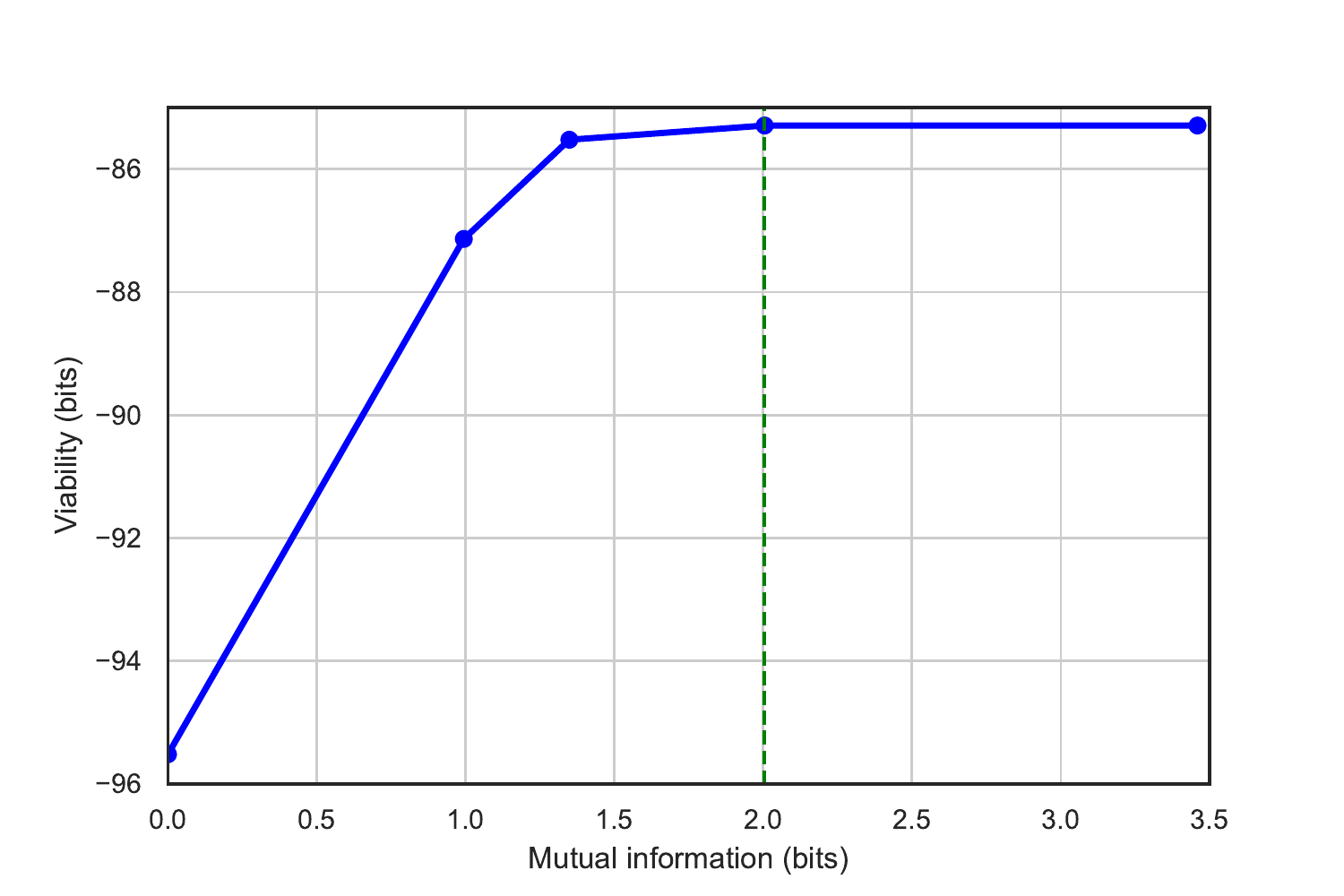}
    \caption{Upper bound on the viability of UE with respect to available information.}
    \label{fig:viability_2}
\end{subfigure}\hfill%
\caption{Viability and transfer entropy in UE handovers.} 
\end{figure*}

\begin{figure*}[t!]
\centering
\begin{subfigure}{1.8\columnwidth}
    \includegraphics[width=6.5in]{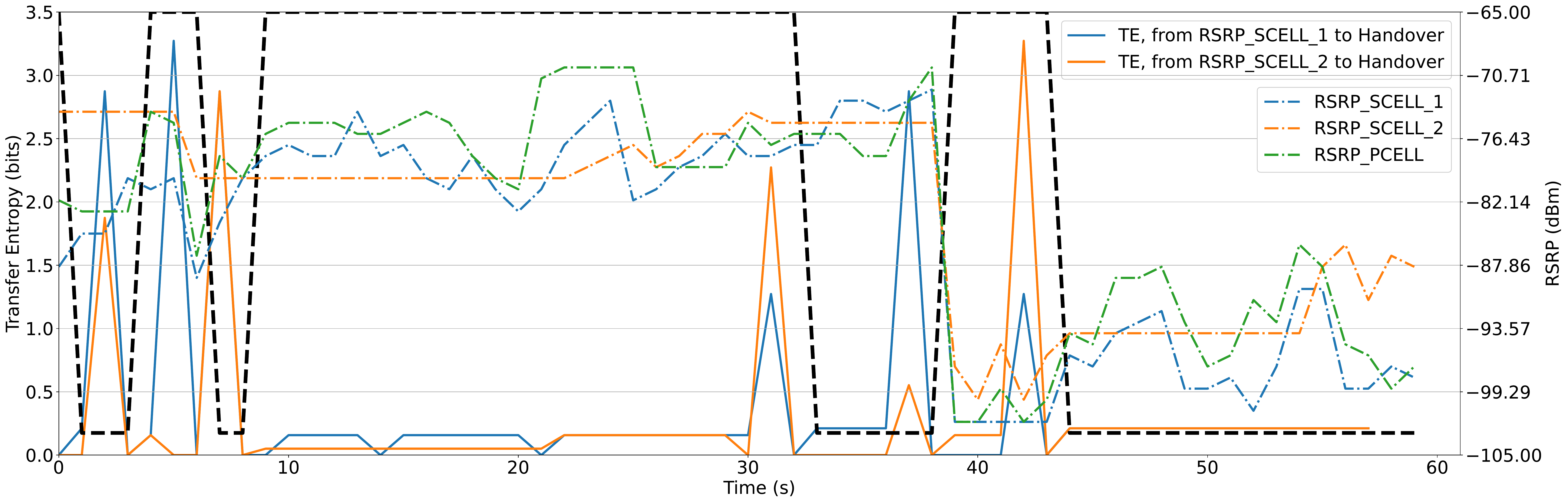}\vspace{-2mm}
    \caption{Relationship between transfer entropy, handover decisions, and RSRP values.  The dashed black line represents the instances of handover from the primary to one of the two secondary cells.}\vspace{2mm}
    \label{fig:use_1}
\end{subfigure}
\begin{subfigure}{1.3\columnwidth}
   \vspace{2mm} \includegraphics[width=4.2in]{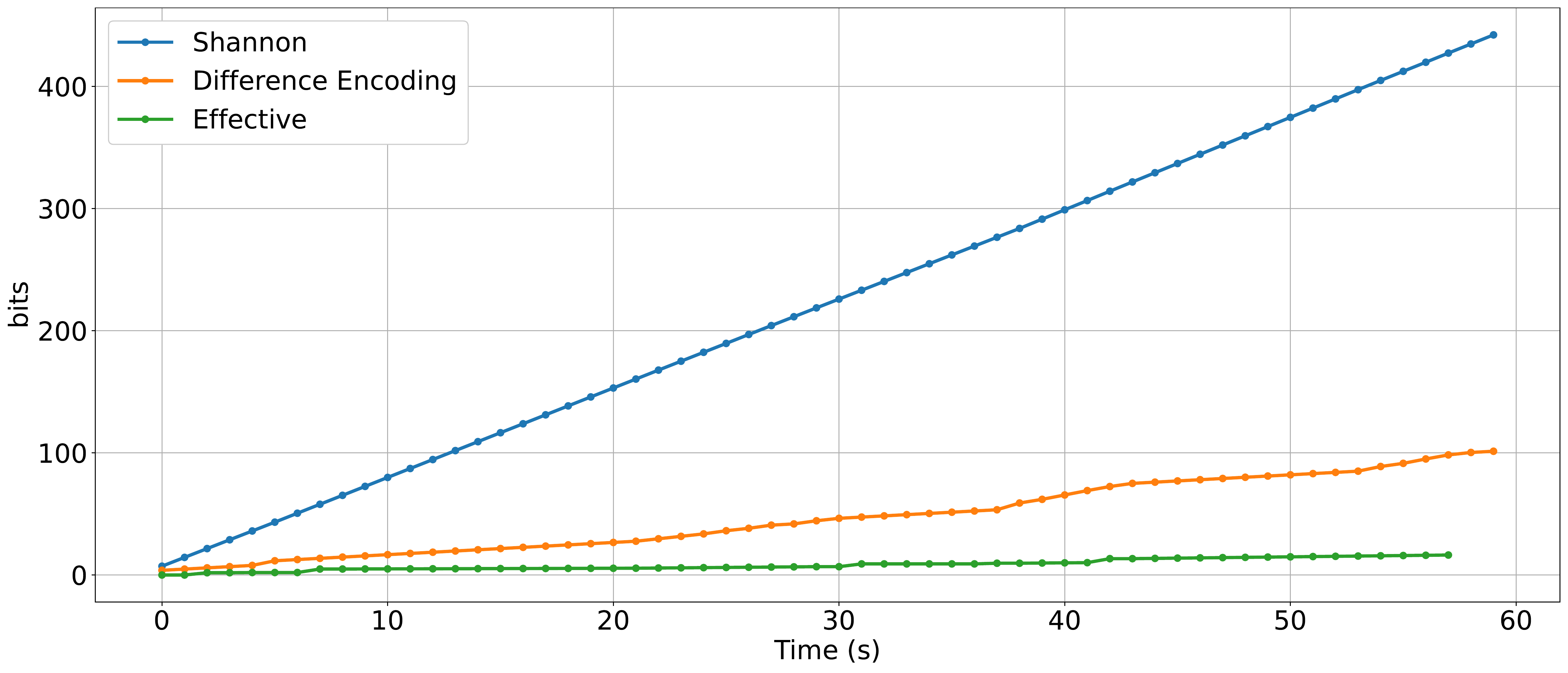}\vspace{-2mm}
    \caption{The cumulative sum of bits transferred for handover mechanism with effective and Shannon communications for the data given in Fig.~\ref{fig:use_1}.}
    \label{fig:use_2}
    \end{subfigure}\hfill%
\caption{Showcasing the data rate for effective information transfer.} 
 \vspace{-3mm}
\end{figure*} 

We first study a fictitious scenario where an edge device measures the Reference Signal Received Power (RSRP) and reports it to the UE. RSRP represents the average power received from a single reference signal, typically varying between -44 dBm and -140 dBm. The message may contain the raw measurement or a function of it, and the UE takes action according to its interpretation of this message. Fig.~\ref{fig:viability_1} illustrates the UE's viability in this scenario when it moves away from its primary BS towards a secondary BS.  Viability is computed as the negative of Shannon entropy.  We assume that time is slotted, and handover is made at integer values of time. Initially, the UE is at the best position in its current BS and the maximum viability of $0$ is attained. We assume that the UE should make its handover by time $3$ (green vertical dashed line),  otherwise the call is dropped. When the edge device shares the raw data of RF measurements from the primary and two secondary BSs, it would require transmitting 24bits per second. However, most of this data is irrelevant and \textit{ineffective} for UE's handover as depicted in Fig.~\ref{fig:viability_2}.  The red dashed line corresponds to the viability where no information regarding handover is delivered from the edge to the UE, and the blue solid line refers to the viability with complete raw data.  At the instant of handover, the value of raw data is at its peak as shown by the difference between the two curves.  As the UE moves without handover, the call of the UE is dropped at time $4$, represented by a viability value of $-100$. Figure \ref{fig:viability_2} depicts the viability as a function of mutual information between the UE and the edge device's measurements. The maximum viability is achieved with only 2 bits of information, which corresponds to the length of the binary representation of the time instant of handover, i.e., at time $3$.  The value represents the viability-optimal minimum transfer entropy, i.e., the minimum amount of information transfer needed to achieve the same level of viability as that with all raw data available.
 
Next, we illustrate the viability-optimal transfer entropy using a real network trace collected from a major cellular operator.  The trace data is collected by a drive-test vehicle over a predetermined route in a predominantly dense urban area with extensive attenuation and shadowing. The data set contains one entry per RF value (e.g., RSRP) and an event (e.g., session start, stall, handover attempt, handover success, etc.). The test UE uses the A2-A4-RSRQ and A3-RSRP handover algorithms, both of which rely on the RSRP.  Figure \ref{fig:use_1} contains a snapshot of 60 seconds of the data set.  The dashed line represents a handover from the primary (PCELL) to one of the two secondary cells (SCELL\_i). During this interval, there are seven handovers.  We plot the differences between the RSRPs of the primary and secondary BSs and the transfer entropy between the RSRP  and the handover action. Transfer entropy is calculated using the Python library PyInform.  Note that the transfer entropy between RSRP and the handover decision represents the rate of minimum information flow from the edge  device to the UE to achieve the correct operation, i.e., UE handover. As expected, we observe that TE takes high values only when the UE needs a handover. 

The choice of transmitted data should achieve a tradeoff between minimizing the number of transferred bits and ensuring the viability of the overall system.  Accordingly, the effective communication will likely have a rate that is less than the Shannon rate but higher than the viability-optimal minimum transfer entropy.  To illustrate this point, we consider an {\em encoding} that relies on transmitting the difference between two subsequent measurements after an initial transmission of the actual measurement.  The source and destination nodes should learn to communicate with this language before the operation, since otherwise, the receiver will not interpret the message correctly.  As Figure \ref{fig:use_2} illustrates, the rate of such encoding is lower than the rate when the raw measurements are transferred (Shannon), but higher than the lowest possible rate defined by the transfer entropy.  It remains an interesting research direction to learn to ``code'' data for effective communication based on the actions of the agents.

\section{Conclusion}
\noindent This study provides  important
new key insights into effective communications defined by Shannon and Weaver in \cite{shannon_weaver_1949}. We studied their requirements and discussed their potential in the face of future 6G applications. We observed that the context-specific accumulation of valuable information could help prevent having to transfer extremely high bit rates. By leveraging deep reinforcement learning architectures, it is possible to identify the relevant, valuable information that can help the receiver maximize its viability, represented by the application experience. We identified the enabling architectures and technologies and discussed the practical and policy challenges to incorporating effective communications on the Internet. Analyzing real-world data, we showcased how much reduction in bit rate is possible in the ideal case of effective communications. The architecture proposed in this paper is far from a complete solution for semantic communication. While we present a partial picture, we hope our discussion will spur interest and further investigations in this area.

\ifCLASSOPTIONcaptionsoff
  \newpage
\fi

\end{document}